# Ideal type-II Weyl points in topological circuits

Rujiang Li[1,2†], Bo Lv[3†], Huibin Tao[4], Jinhui Shi[3], Yidong Chong[5,6*], Baile Zhang[5,6*] and Hongsheng Chen[1,2*]

[1]State Key Laboratory of Modern Optical Instrumentation and The Electromagnetics Academy at Zhejiang University, Zhejiang University, Hangzhou, China.

[2]Key Laboratory of Micro-Nano Electronics and Smart System of Zhejiang Province, College of Information Science and Electronic Engineering, Zhejiang University, Hangzhou, China.

[3]Key Laboratory of In-Fiber Integrated Optics of Ministry of Education, College of Science, Harbin Engineering University, Harbin, China.

[4]School of Software Engineering, Xi'an Jiaotong University, Xi'an, China.

[5]Division of Physics and Applied Physics, School of Physical and Mathematical Sciences, Nanyang Technological University, Singapore 637371, Singapore.

[6]Centre for Disruptive Photonic Technologies, Nanyang Technological University, Singapore 637371, Singapore.

[†]These authors contributed equally to this work.

*Correspondence to:

yidong@ntu.edu.sg, blzhang@ntu.edu.sg, hansomchen@zju.edu.cn

**Weyl points (WPs), as nodal degenerate points in three-dimensional (3D) momentum space, are ideal if they are symmetry-related, well-separated, residing at the same energy and far from the nontopological bands. Although type-II WPs show some unique features compared with type-I counterparts, ideal type-II WPs have not yet been reported due to the lack of an ideal Weyl system with enough flexibility to tilt the dispersion bands. By stacking two-dimensional (2D) layers of inductor-capacitor ($LC$) resonator dimers with the breaking of parity inversion symmetry, here we experimentally realize the topological circuits with only the topological bands and observe a minimal number of four ideal type-II WPs. Two**

**hallmark features of type-II WPs: a strongly tilted band structure with two group velocities having the same sign near type-II WPs and the topological surface states in an incomplete bandgap have been demonstrated. Our results establish an ideal system to the further study of Weyl physics and provide a new perspective to the access of topological matters.**

Weyl point (WP) is a nodal degenerate point where two linear energy spectra intersect in 3D momentum space (*1*). In a 3D crystal, if all the WPs are symmetry-related, residing at the same energy with a large momentum separation and devoid of nontopological bands in a sufficiently large energy interval, such a crystal is called as an ideal Weyl system (*2-4*). Due to these features, in an ideal Weyl system it is less challenging to identify the real WPs from the subtle band crossings and to demonstrate the intriguing phenomena such as the topological surface states (*5*). Moreover, ideal WPs are particularly useful in some realistic and innovative device applications (*6*). To date, ideal type-I WPs with symmetric cone spectra have been observed in semimetals (*2,7*), and also in artificial photonic crystal structures utilizing the flexibility and diversity of classical systems (*4*).

However, as another important constituent of ideal WPs, ideal type-II WPs have not yet been reported although type-II WPs can be transited from type-I counterparts by increasing the kinetic component in the Weyl Hamiltonian (*8*). In contrast with the well-studied type-I WPs (*9-13*), it was not until 2015 that type-II WPs were theoretically proposed (*14*) and then experimentally confirmed in condensed matter (*15-17*), photonic (*18-20*) and acoustic systems (*21*). Type-II WPs show some unique features which include strongly tilted cone spectra, two group velocities having the same sign near type-II WPs and topological surface states located in an incomplete bandgap (*14,22*). Besides, type-II WPs can be used to explore the fascinating analogues in astrophysics such as the Hawking radiation and gravitational lensing (*23,24*). Unfortunately, the realization of ideal type-II WPs is hindered by the lack of an ideal Weyl system with enough flexibility to tilt the dispersion bands (*8*). To the best of our knowledge, ideal type-II WPs were only theoretically predicted in a few condensed matter systems without experimental validation (*3,25,26*).

Here we report on the realization of a minimal number of ideal type-II WPs in topological circuits that reside at the same energy with a large momentum separation. Using the site-resolved transmission measurement and mapping out the band structures, we confirm the existence of a strongly tilted band structure with two group velocities having the same sign near type-II WPs and the topological surface states in an incomplete bandgap, which are two hallmark features of type-II WPs. Unlike the previous proposals (*3,25,26*), our design utilizes the macroscopic circuit system. Its connectivity can be literally wired in an arbitrary manner with arbitrary numbers of connections per node and long-range connections, and its hopping is independent on the distance between the nodes (*6*). This flexible and controllable connectivity and the distance independent hopping allow an easy fabrication of a two-band topological circuit without any nontopological bands. Under *P* breaking and *T* preserving, we obtain the minimal number of four ideal type-II WPs, which reside at the same energy with a large separation.

Our topological circuit exhibiting ideal type-II WPs is constructed by stacking 2D layers of *LC* resonator dimers with the breaking of *P* (*22*). As depicted in Fig. 1a, the white and black nodes are wired to two types of grounded parallel *LC* resonators, respectively, and the nearest nodes are wired by the coupling capacitors in the *x*, *y* and *z* directions. The inductors in all the resonators are identical. For an isolated layer in the *x-y* plane with the identical grounded capacitors $C_a = C_b$ and the identical coupling capacitors $C_1 = C_2 = C_3$, its two bands degenerate at the square 2D BZ boundary and there are quadratic degeneracies at the corners (*27*). By stacking identical layers along the *z* direction with $C_4 = C_5$, the band degeneracy points (BDPs) form a square tube with its rotational axis along the $k_z$ direction (supplementary information sections 3 and 4). In order to break $P$ of the circuit, we break the mirror symmetry in the $x$ direction $M_x := x \rightarrow -x$ by letting $C_1 \neq C_2$, which leads to the splitting of the square degeneracy in the $k_x$-$k_y$ plane (*22,27*). In the three-dimensional (3D) BZ, the BDPs form a pair of lines along the $k_z$ direction with $k_x = 0$ and $k_y$ positions being determined by the sum of $C_1$ and $C_2$, and they have linear dispersion in the $k_x$ and $k_y$ directions. Here we let $C_1 + C_2 = C_3$ and the degenerate lines are projected to $(0, \pm 2/3)\pi/a$, where $a$ is the spacing between the nearest resonator nodes. To further realize the isolated BDPs in the BZ, we break the line degeneracy by letting $C_4 \neq C_5$. In this case, two BDPs with quadratic dispersion in the $k_z$ direction are located at $(0, \pm 2/3, 0)\pi/a$. In order to move the BDPs to the positions that have linear dispersion, we use resonators with two different resonant frequencies by letting $C_a \neq C_b$ and specially we let $C_a - C_b = -2(C_4 - C_5)$. Then the original two BDPs split into four points at $(0, \pm 2/3, \pm 1/2)\pi/a$, which are the type-II WPs.

Under $P$ breaking and $T$ preserving, the minimum number of type-II WPs is four (*6*). As shown in Fig. 1**b**, the four type-II WPs are distributed in momentum space with a large separation, where the first BZ is enclosed by dashed lines. Type-II WPs exist in pairs with opposite chirality and each point acts as either a source (red spheres) or sink (blue spheres) of Berry curvature (supplementary information section 5). This topological circuit has two mirror symmetries $M_y := y \rightarrow -y$ and $M_z := z \rightarrow -z$, and $T$. Considering that mirror symmetry requires a WP located at **k** to create another WP at **-k** with the opposite chirality and $T$ requires another WP at **-k** to have the same chirality (*6*), the four type-II WPs satisfy the constrains imposed by the symmetry i.e. they are symmetry-related. Notably all these type-II WPs reside at the same energy $f = 104.0$ kHz under the realistic circuit parameters mentioned in the caption of Fig. 1. Besides, the capacitive coupling from the identical resonators in the $z$ direction breaks the Lorentz symmetry, which leads to a strongly tilted band structure in the $k_x$-$k_z$ plane with $k_y = \pm 2\pi/3a$, as shown in Fig. 1**c**. From the above discussion, our topological circuit is an ideal type-II Weyl system since the minimal number of four type-II WPs are symmetry-related, well-separated and residing at the same energy, and there are no nontopological bands.

In order to verify the existence of ideal type-II WPs, we experimentally implement two topological circuits using stacked circuit boards, where one is periodic along the *x*, *y* and *z* directions and another one is finite along the $p=(x-y)/\sqrt{2}$ direction and periodic along the $q=(x+y)/\sqrt{2}$ and *z* directions (supplementary information section 1). Since a periodic boundary can mimic an infinite boundary (*27*), the two topological circuits are equivalent to two theoretical circuit lattices, respectively, where one is infinite along the three directions and another one is finite along the *p* direction and infinite along the other directions. Experimentally we excite and probe the resonator nodes of the topological circuits using a network analyzer. Then the band structures in different directions of the BZ are mapped out by applying a Fourier transform to the complex transmission coefficients in real space to validate the existence of ideal type-II WPs (supplementary information sections 2 and 6).

One exciting signature of type-II WPs is a strongly tilted band structure with two group velocities having the same sign near type-II WPs. As shown in Figs. 2**a-b**, we fabricate the topological circuit with the periodic boundaries along the three directions. The assembled 3D circuit structure contains eight layers and it has a size of 23.5 cm × 31.0 cm × 13.5 cm. Experimentally, bulk states are excited and probed by a site-resolved measurement, and the one-dimensional (1D) band structures are shown in Figs. 2**c-e**. The corresponding theoretical band structures calculated from the equivalent circuit lattice which is infinite along the three directions are shown in in Figs. 2**f-h**. We sweep $k_x$ at $(k_y, k_z)$ = (±2/3, ±1/2)π/*a* in Figs. 2**c** and **f**, and get one degenerate point at $k_x$ = 0. Similarly, by sweeping $k_y$ at $(k_x, k_z)$ = (0, ±1/2)π/*a* in Figs. 2**d** and **g**, we get two degenerate points at $k_y$ = ±2π/3*a*. In Fig. 2**e** and **h**, there are two degenerate points located at $k_z$ = ±π/2*a* by sweeping $k_z$ at $(k_x, k_y)$ = (0, ±2/3)π/*a*. For each point, the two strongly tilted bands imply that their group velocities have the same sign, which is a signature of type-II WPs. Specially, according to the **k·p** model, the band structure has linear dispersion at these ideal type-II WPs (supplementary information section 7).

Another exciting signature of type-II WPs is the existence of topological surface states in an incomplete bandgap. As shown in Fig. 3**a-b**, we fabricate the topological circuit with a finite boundary along the *p* direction and the periodic boundaries along the other directions. The assembled circuit structure also contains eight layers and it has a size of 24.0 cm × 40.0 cm × 13.5 cm. In order to excite the surface states, we excite the resonator nodes from the two boundaries normal to the -*p* and +*p* directions, respectively. The experimental band structures are shown in Figs. 3**c-d**. Similarly, the corresponding theoretical band structures calculated from the equivalent circuit lattice which has 2.5 unit cells along the *p* direction and infinite along the other directions are shown in Figs. 3**e-f**. In Fig. 3**e**, the red curves in the dispersion spectrum with $k_z = \pm\pi/2a$ which is confirmed by Fig. 3**c** denote two nondegenerate bands, where $k_q = (k_x + k_y)/\sqrt{2}$. When the momentum is lying on the bands (red curves) with $|k_q| \geq 2\pi/3a$, there are two distinct surface states which are located in an incomplete bandgap with the same sign of group velocities (*22*), as shown by the red and blue curves in Fig. 3**f** with $k_q = \pm\pi/a$, and the six bright spots around *f* = 104.0 kHz. In order to further demonstrate the topological surface states, in the inset of Fig. 3**d** we show the transmission distributions on a part of the circuit, when the circuit is excited at *f* = 104.0 kHz from the two boundaries, respectively. In both cases, the transmission peaks probed at the two opposite boundaries validate the existence of the topological surface states with $(k_q, k_z) = (\pm\pi/a, \pm\pi/2a)$.

Based on the two implemented topological circuits, we have demonstrated a strongly tilted band structure with two group velocities having the same sign near type-II WPs and the topological surface states in an incomplete bandgap. The two signatures validate the existence of a minimum number of four ideal type-II WPs that reside at the same energy with a large momentum separation and imply that our circuit system is an ideal type-II Weyl system. Compared with other 3D systems, periodic boundaries are easier to be implemented in circuit systems, which makes it feasible to construct band structures using compact circuits. Besides, the performance of the circuit system such as the frequency of type-II WPs and the group velocities near type-II WPs can be adjustable (supplementary information section 8). These advantages imply that our topological circuits provide a clean and adjustable platform to observe ideal type-II WPs.

The designed ideal type-II Weyl system presented here opens the door to the realization of new topological phases of matter and provides a prototype platform for realistic device applications. The flexibility and controllability of our circuit system are particularly beneficial for the nonlinear (*28*), higher-order (*29*) and higher-dimensional (*30*) topological systems. Although our demonstration was carried out between 80-130 kHz, the design principle should be generalizable to other frequency regimes such as microwave frequencies using microstrips (*31*) and optical frequencies using subwavelength nanoparticles (*32*). Furthermore, the building blocks of our circuit, *LC* resonators, can be regarded as atoms of a periodic lattice, i.e. a crystal, and the capacitive coupling corresponds to the bonding between these atoms. A similar lattice design may be applied to other classical systems such as photonic and acoustic systems to observe ideal type-II WPs.

**Data availability.** The data that support the plots within this paper and other finding of this study are available from the corresponding author upon reasonable request.

References
1. Armitage, N. P., Mele, E. J. & Vishwanath, A. Weyl and Dirac semimetals in three dimensional solids. Rev. Mod. Phys. **90**, 015001 (2018).

4D topological insulator by electric circuits. arXiv: 1906.00883 (2019).

31. Li, Y. et al. Topological *LC*-circuits based on microstrips and observation of electromagnetic modes with orbital angular momentum. Nat. Commun. **9**, 4598 (2018).

32. Engheta. N. Circuits with Light at Nanoscales: Optical Nanocircuits Inspired by Metamaterials. Science **317**, 1698-1702 (2007).

<>**Acknowledgements**

We thank B. Yan and Y. Liu for helpful discussions. R.L. and H.C. was sponsored by the National Natural Science Foundation of China under Grant Nos. 61625502, 61574127, 61601408, 61775193 and 11704332, the ZJNSF under Grant No. LY17F010008, the Top-Notch Young Talents Program of China, the Fundamental Research Funds for the Central Universities under Grant No. 2017XZZX008-06, and the Innovation Joint Research Center for Cyber-Physical-Society System. B.L. was sponsored by the Fundamental Research Funds for the Central Universities under Grant No. 3072019CFJ2504 and the National Natural Science Foundation of China under Grant No. 6190010226. J.S. was sponsored by the National Natural Science Foundation of China under Grant Nos. 91750107 and 61875044. Y.C. and B.Z. acknowledge the support of Singapore Ministry of Education under grant numbers MOE2015-T2-1-070, MOE2015-T2-2-008, MOE2016-T3-1-006 and Tier 1 RG174/16 (S).

**Author contributions**

R.L. and B.L. conceived the idea, carried out the calculations and processed the experimental data. R.L., B.L. and H.T. designed the experimental setup and made the measurements. Y.C., B.Z. and H.C. supervised the project. All authors contributed to the discussion and writing of the manuscript.

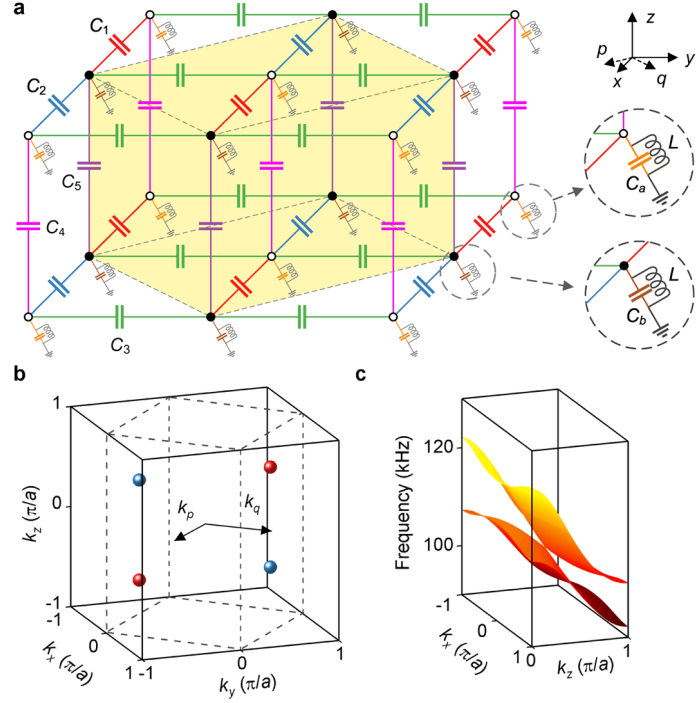

**Fig. 1 Topological circuit with ideal type-II WPs. a,** Schematic of the circuit, where the white and black nodes are wired to the grounded parallel $LC$ resonators with two different resonant frequencies, respectively, and the nearest nodes are wired by the coupling capacitors in the $x$, $y$ and $z$ directions. The coupling capacitors are $C_1$ = 18 pF, $C_2$ = 82 pF, $C_3$ = 100 pF, $C_4$ = 270 pF and $C_5$ = 120 pF, the grounded capacitors are $C_a$ = 1.5 nF and $C_b$ = 1.8 nF, and the inductors are $L$ = 1 mH. The yellow cube denotes a unit cell. The $p$ and $q$ directions are defined as $p = (x-y)/\sqrt{2}$ and $q = (x+y)/\sqrt{2}$, respectively. **b,** The topological circuit creates a minimal number of four ideal type-II WPs that are located at $(0, \pm 2/3, \pm 1/2)\pi/a$ in momentum space and residing at the energy $f$ = 104.0 kHz, where the 3D first BZ is enclosed by dashed lines and $a$ is the spacing between the nearest resonator nodes. The $k_p$ and $k_q$ directions are defined as $k_p = (k_x - k_y)/\sqrt{2}$ and $k_q = (k_x + k_y)/\sqrt{2}$, respectively. The type-II WPs exist in pairs with opposite chirality and each point acts as either a source (red spheres) or sink (blue spheres) of Berry curvature. **c,** A strongly tilted 2D band structure for the circuit in the $k_x$-$k_z$ plane with $k_y = \pm 2\pi/3a$.

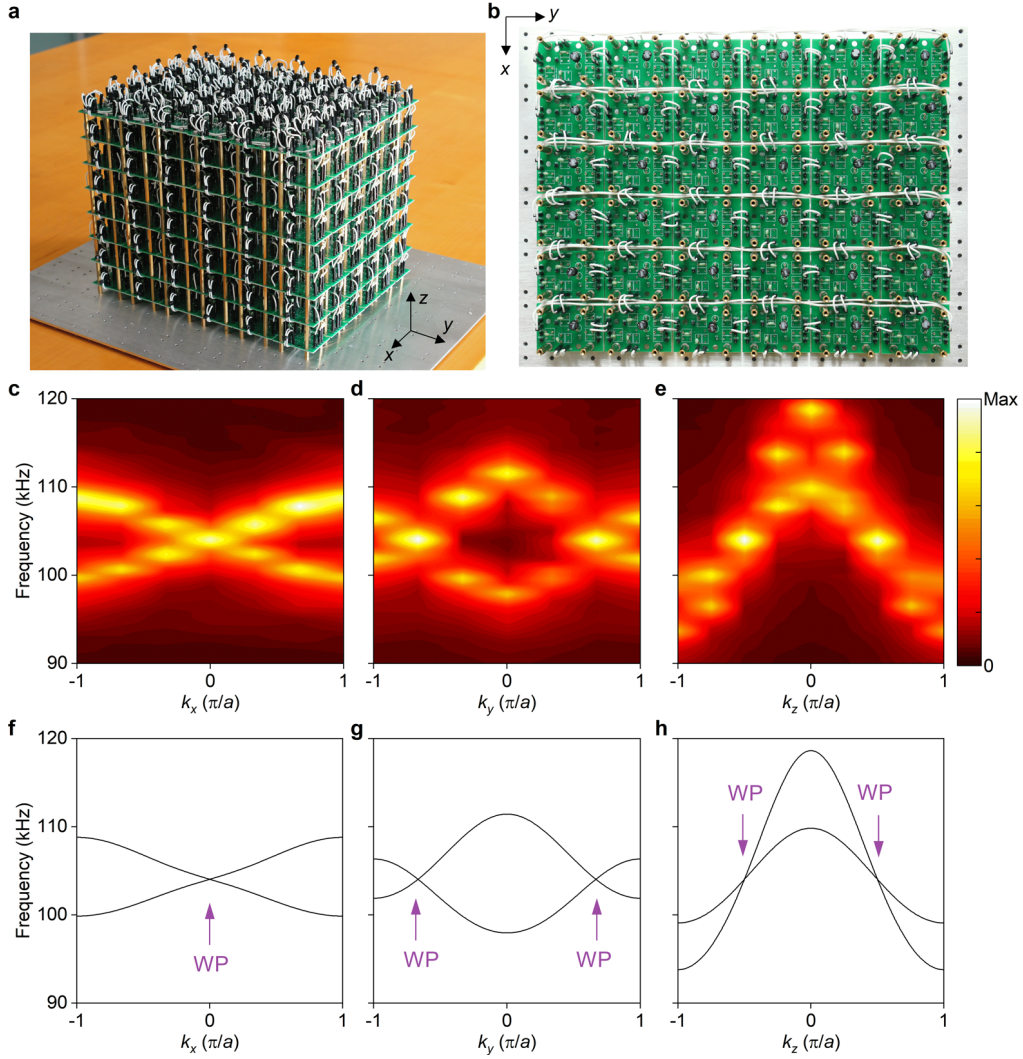

**Fig. 2 Observation of a strongly tilted band structure. a-b,** 3D schematic (**a**) and bottom layer (**b**) of the topological circuit which is periodic along the *x*, *y* and *z* directions. **c-e,** 1D experimental band structures by sweeping $k_x$ at ($k_y$, $k_z$) = (±2/3, ±1/2)π/*a*, $k_y$ at ($k_x$, $k_z$) = (0, ±1/2)π/*a* and $k_z$ at ($k_x$, $k_y$) = (0, ±2/3)π/*a*, respectively. **f-h,** The corresponding theoretical band structures calculated from the equivalent circuit lattice which is infinite along the three directions. Four ideal type-II WPs are located at (0, ±2/3, ±1/2)π/*a* and residing at the same energy *f* = 104.0 kHz. In **e** and **h**, the two strongly tilted bands imply that their group velocities have the same sign, which is a signature of type-II WPs.

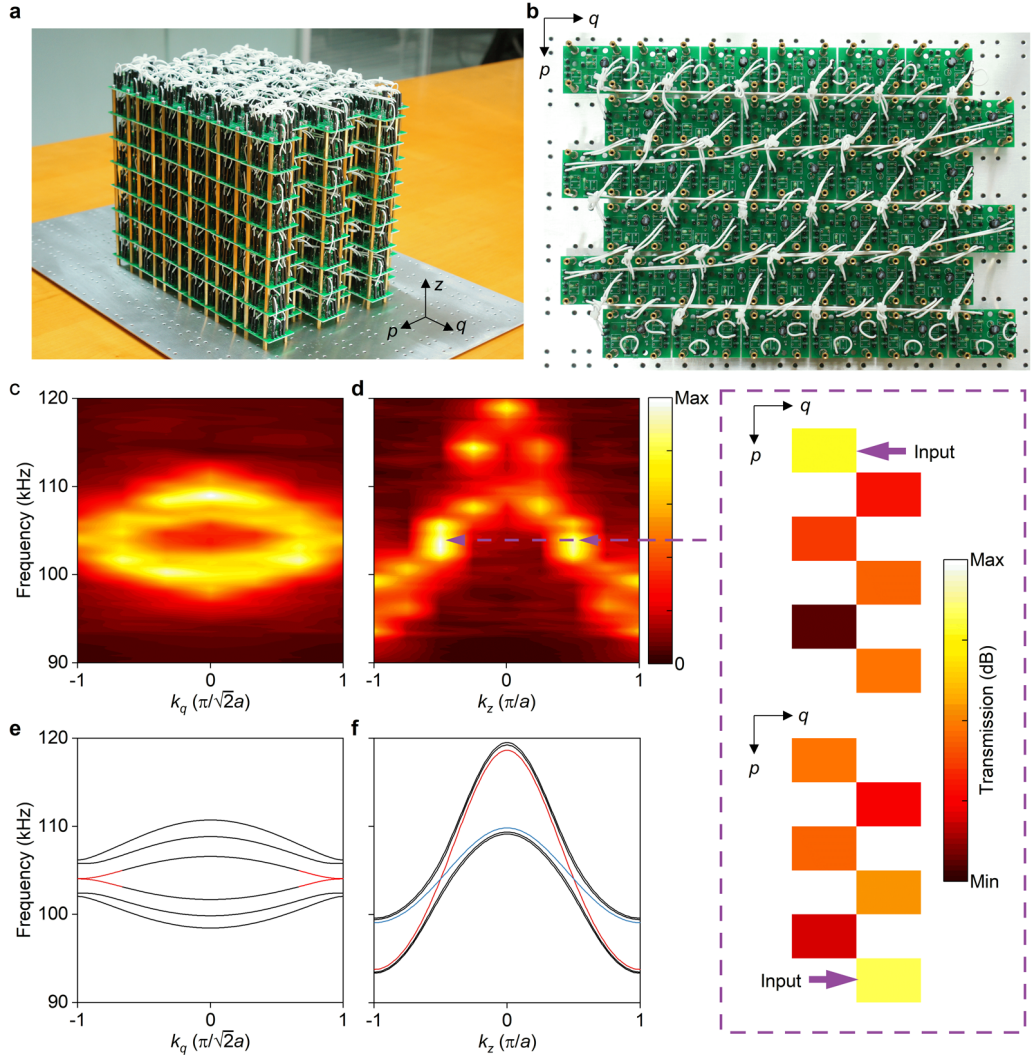

**Fig. 3 Observation of topological surface states in an incomplete bandgap. a-b,** 3D schematic (**a**) and bottom layer (**b**) of the topological circuit which is finite along the $p$ direction and periodic along the $q$ and $z$ directions. **c-d,** Experimental band structures along the $k_q$ direction with $k_z = \pm\pi/2a$ and along the $k_z$ direction with $k_q = \pm\pi/a$, respectively. **e-f,** The corresponding theoretical band structures. When the momentum is lying on the red curves in **e** with $|k_q| \geq 2\pi/3a$, there are two distinct topological surface states in an incomplete bandgap, as indicated by the red and blue curves in **f**. The black curves in **e-f** correspond to the bulk states. The inset of **d** shows the transmission distributions of the surface states with $(k_q, k_z) = (\pm\pi/a, \pm\pi/2a)$ when the circuit is excited at $f = 104.0$ kHz from the two boundaries, respectively. In both cases, the transmission peaks probed at the two boundaries validate the existence of surface states, which is another signature of type-II WPs.